\documentclass[%
 aip,
 superscriptaddress,
 amsmath,amssymb,
 reprint,%
]{revtex4-1}

\usepackage{tikz}
\usepackage[ruled,vlined,linesnumbered]{algorithm2e}  
\usepackage{graphicx}
\usepackage{bm}
\usepackage{color}
\usepackage{cleveref}

\begin{document}

\title{The topology of a chaotic attractor in the Kuramoto-Sivashinsky equation}

\author{Marie Abadie}\email{marie.abadie@uni.lu}
\address{
Department of Mathematics, University of Luxembourg, 6, Av. de la Fonte, 4364 Esch-sur-Alzette, Luxembourg}

\author{Pierre Beck}\email{pierre.beck@epfl.ch}
\address{
Emergent Complexity in Physical Systems Laboratory (ECPS), \'Ecole Polytechnique F\'ed\'erale de Lausanne, 1015 Lausanne, Switzerland
}

\author{Jeremy P. Parker}\email{jparker002@dundee.ac.uk}

\address{
Division of Mathematics, University of Dundee, Dundee DD1 4HN, United Kingdom
}

\author{Tobias M. Schneider}\email{tobias.schneider@epfl.ch}
\address{
Emergent Complexity in Physical Systems Laboratory (ECPS), \'Ecole Polytechnique F\'ed\'erale de Lausanne, 1015 Lausanne, Switzerland
}

\begin{abstract}
    The Birman-Williams theorem gives a connection between the collection of unstable periodic orbits (UPOs) contained within a chaotic attractor and the topology of that attractor, for three-dimensional systems. In certain cases, the fractal dimension of a chaotic attractor in a partial differential equation (PDE) is less than three, even though that attractor is embedded within an infinite-dimensional space. Here we study the Kuramoto-Sivashinsky PDE at the onset of chaos. We use two different dimensionality-reduction techniques -- proper orthogonal decomposition and an autoencoder neural network -- to find two different mappings of the chaotic attractor into three dimensions. By finding the image of the attractor's UPOs in these reduced spaces and examining their linking numbers, we construct templates for the branched manifold which encodes the topological properties of the attractor. The templates obtained using two different dimensionality reduction methods are equivalent. The organization of the periodic orbits is identical and consistent symbolic sequences for low-period UPOs are derived. While this is not a formal mathematical proof, this agreement is strong evidence that the dimensional reduction is robust, in this case, and that an accurate topological characterization of the chaotic attractor of the chaotic PDE has been achieved.
\end{abstract}

\maketitle

\begin{quotation}
The topological properties of a dynamical system describe its behaviour in a way which is independent of the choice of coordinates used. In particular, the pattern of stretching and folding characteristic of dissipative chaos is encoded by the topological properties of the periodic orbits and the branched manifold on which they lie. It is well-known that the long-time behaviour of many dissipative partial differential equations is finite-dimensional, but exploiting this fact to determine the topology of the dynamics in practice has not been widely studied. Here we present a procedure to do so in the case that the underlying dynamics are three-dimensional, and apply it to the Kuramoto-Sivashinsky equation.
\end{quotation}

\section{Introduction}


The importance of unstable periodic orbits (UPOs) to the understanding of chaotic dynamics has long been appreciated. It is believed, and in some cases proven, that UPOs are dense within strange attractors. Modern computational methods enable the detection and convergence of large numbers of such UPOs in high-dimensional discretizations of dynamical systems governed by partial differential equations (PDEs), including the Navier-Stokes equations from fluid dynamics \citep{budanur2017relative, parker2022variational, page2024exact, mccormack2024multi}.
Especially in the context of fluid turbulence, UPOs carry the hope that we may be able to describe statistical properties of the turbulent flow as weighted ensemble averages over these UPOs. This envisioned form of `periodic orbit theory' would transfer well-studied ergodic theory from low-dimensional chaotic ordinary differential equation (ODE) systems \citep{ChaosBook, lan2010cycle, eckhardt1994periodic} to PDEs. 

One alternative approach to use periodic orbits for understanding the behaviour of chaotic attractors is through ideas from topology. In three dimensions, UPOs, which are closed loops in state space, form knots, and pairs of UPOs form links.
Such knots and links possess topological invariants, which can be used to analyse the topological structure of the underlying chaotic attractor.
In particular, the Birman-Williams theorem \citep{Birman1983KnottedPO, birman1983knotted} shows that a chaotic attractor in a three-dimensional system can be identified with a two-dimensional branched manifold, a so-called `template', on which all periodic orbits lie. The structure of this branched manifold -- the number of branches and how they link together -- can be deduced from the linking numbers of the periodic orbits. 
This formalises the observation that chaotic attractors in canonical examples such as the Lorenz and R\"ossler systems seem to be very thin, almost two-dimensional structures, and the templates found for these systems match the structures visible in a long chaotic trajectory in the usual coordinates.
The method of determining templates for all periodic orbits has been successfully applied to many three-dimensional ODE systems, including ones for which the topology is not obvious from observing trajectories \citep{Birman1983KnottedPO, Letellier1995UnstablePO,letellier2012required}. 

In four and higher dimensions, all UPOs are however equivalent (i.e. ambient isotopic) to the unknot, and thus the specific methods discussed here cannot be directly applied.
The solution spaces of PDEs are infinite dimensional. In practice, when solving PDE dynamical systems numerically, this infinite dimensional space is discretized to be finite-dimensional, but the number of dimensions is still on the order of at least tens, if not much higher, and so the ideas of Birman and Williams are not directly applicable (although more modern approaches exist which can be applied to dimensions higher than three, as discussed in \cref{sec:conclusion}).
Despite the high-dimensional solution spaces, the long-time dynamics of many systems often collapses onto an attracting set of much lower dimension.

The dissipative PDE systems that arise in fluid dynamics are typically associated with a parameter, such as the Rayleigh number or Reynolds number, whose increase is correlated with an increase in the complexity of the dynamics. When such a parameter is sufficiently high the system is fully turbulent, as characterised by a large number of positive Lyapunov exponents. In this case, the global attractor is a relatively high- (but still finite-) dimensional fractal structure in state space. In contrast, when the parameter is sufficiently small, the system collapses onto a single attracting fixed point. Between these regimes, it is often possible to find chaotic dynamics with only one positive Lyapunov exponent, which indicates a strange attractor with less than three dimensions (in the absence of symmetries), embedded within the infinite-dimensional state space of the PDE. If it were possible to construct an embedding of this low dimensional attractor into three-dimensional space, this would preserve the topological properties, and we would be able to use the ideas of Birman and Williams to find a template for the system.

Since the topological properties of dynamical systems are independent of the choice of coordinates used for parameterising state space, they provide a way to compare different representations of the same or similar systems. This is particularly relevant to PDE systems, in which the dynamics are necessarily discretized in order to be solved on a computer. Different choices of discretization (for example, finite differences or Galerkin methods) may lead to very different state space geometries, but the topology must be equivalent if the discretizations are accurate. Furthermore, it is often desirable to find a `reduced-order-model', a small set of ordinary differential equations approximating the dynamics of a PDE system, either to increase computational efficiency or to aid the interpretation of physical mechanisms underlying the dynamics. 
One way to validate the accuracy of such a model would be to check that the model preserves the topology of the dynamics of the full system.
However, in order for this we need a method to distill the topology of a chaotic PDE, which does not currently exist.

Previous authors have attempted to find reduced-order models with the same dynamics as infinite dimensional systems, for example by using global modeling to model the dynamics with a three-dimensional ODE, from which a template can be derived \citep{ghosh2017tumor}. Our approach is instead to consider accurate periodic trajectories of the full PDE system, and then to embed these trajectories in a three-dimensional reduced space.

Reducing the state space of a high-dimensional dynamical system to an accurate but approximate low-dimensional representation is an important and active area of research. Traditional methods, such as proper orthogonal decomposition \citep{lu2019review} (POD) and dynamic mode decomposition\citep{schmid2022dynamic}, use numerical linear algebra tools to determine the most important `modes' from a timeseries. This is often successful, although these methods have a fundamental limitation when the dynamics and state-space geometry are strongly nonlinear. For these reasons, researchers have applied the more flexible and nonlinear methods of deep learning to dimensionality reduction. In this paper we use POD as well as the more modern approach of an autoencoder, a pair of deep neural networks which are trained to find an invertible (as close as possible) mapping from the full state space to a low-dimensional `latent' space and back. Due to their nonlinearity, autoencoders have been applied as a dimensionality reduction tool to a wide variety of problems in nonlinear dynamics, including chaotic systems. 
Questions on the interpretability of the latent space, such as the minimum number of latent dimensions and the physical meaningfulness of latent coordinates have also been studied. \citet{Linot2022} investigate how the errors of these networks vary depending on how close the latent dimension is to the manifold dimension of the chaotic attractor. \citet{PageBrennerKerswell2020} showed that by applying deep convolutional autoencoders to 2D Navier-Stokes, they are able to identify meaningful low-dimensional representations of two-dimensional turbulence. The apparent success of autoencoders at finding a low-dimensional representations for such complicated systems indicates the potential that the methods presented in this paper may be applicable to higher-dimensional systems including chaotic fluid flows.

The idea that autoencoders can in some cases preserve the topology of a chaotic attractor is supported by the studies of \citet{Mindlin1} and \citet{Mindlin2}, who applied autoencoders to timeseries from synthetic and experimental data. Unlike in the present work, the aim there was to determine whether the known topological properties are preserved so that the autoencoders give an embedding. Indeed, both of these papers showed that this is not always the case, and so autoencoders should be used with caution.

We consider the Kuramoto-Sivashinsky equation (KSE), a dissipative PDE in one spatial dimension, which is frequently used as a model for the Navier-Stokes equations and other more complicated PDE systems. In this system, the domain size $L$ dictates the complexity of the dynamics. When constrained to its antisymmetric subspace, stable chaotic dynamics is first observed at $L\approx18$, whose existence was proven by \citet{WILCZAK20208509}.
This chaotic attractor is mapped into a three-dimensional space using both an autoencoder and also a traditional POD analysis. Using these two mappings to $\mathbb{R}^3$, we can compute linking numbers for the known periodic orbits and construct a template for them. If both mappings preserve the topology, as hypothesized, the template should be equivalent for the POD and the autoencoder-based mapping.

The topological properties of chaos within the KSE (in a very different parameter regime) have been studied by \citet{siminos2021manifold} using an entirely different and more general technique. In it, the dimensionality reduction technique of manifold learning was applied to Poincar\'e return maps, essentially learning the topology of a discrete-time dynamical system. By examining the linking of periodic orbits, our method gains a further understanding of the topology as a continuous-time system.

The remainder of this paper is laid out as follows. In \cref{sec:kse} the PDE system we study is described. In \cref{sec:reduction} we explain the dimensionality reduction methods and their application to the system and its periodic orbits. In \cref{sec:topology} we give background on the topological approach to periodic orbits, and in \cref{sec:template} the periodic orbits are used to find a template for the system. Finally, concluding remarks are given in \cref{sec:conclusion}.

\section{The Kuramoto-Sivashinsky equation}
\label{sec:kse}

The Kuramoto-Sivashinsky equation (KSE) is a dissipative partial differential equation of one spatial dimension, originally derived to model the dynamics of flame fronts \citep{sivashinsky1977flames}. It is believed to be the simplest PDE which exhibits spatio-temporal chaos \citep{brummitt2009search}, and thus is often used as a model system on which to test methods before they are applied to more complicated PDEs such as the Navier-Stokes equations \citep{christiansen1997spatiotemporal,zoldi1998spatially}. Many different equivalent formulations are found in the literature; we use
\begin{equation}
    \partial_t u + u \partial_x u + \partial_x^2 u + \partial_x^4 u = 0,
    \label{eq:kse}
\end{equation}
subject to periodic boundary conditions in $x$. In this form, the only parameter of the system is the domain size, $L$. An increase in $L$ is generally associated with an increase in the complexity of the flow, as characterised by the number of positive Lyapunov exponents \citep{edson2019lyapunov}.

\Cref{eq:kse} is invariant under the transformation $x\mapsto-x$, $u\mapsto-u$ and therefore admits an invariant subspace of solutions for which $u(-x)=-u(x)$. Following previous studies\citep{cvitanovic2010state, lasagna2018sensitivity, dong2018topological}, we consider the dynamics constrained to this (usually unstable) subspace. This has the advantage of removing the continuous symmetry $x\mapsto x+\Delta$, which otherwise would increase the dimension of any $x$-varying attractor by 1, significantly complicating topological analysis.
In this subspace, the first known chaotic attractor occurs at $L\approx18.05$. \citet{WILCZAK20208509} give a computer assisted proof for the existence of this attractor at $L=2\pi/\sqrt{0.1212}=18.048\dots$, which is the parameter value we use. Indeed, they proved the existence of a symbolic dynamics for the system, which in principle allows the enumeration and computation of a countable infinity of UPOs.
We found, empirically, that this attractor has a very small basin of attraction: it is necessary to use an initial condition very close to the states depicted in \citet{WILCZAK20208509}, otherwise the system rapidly collapses onto a fixed point.

To have any hope of applying methods based on knot theory to such a system, the fractal dimension of the attracting chaotic set must be less than three. An upper bound is given by the Lyapunov dimension \citep{constantin1983global}, which we calculate to be approximately $2.227$ based on the four leading exponents in the Lyapunov spectrum being $0.0149$, $0$, $-0.0658$ and $0.0989$. Note that no positive Lyapunov exponents were found at this parameter value by \citet{edson2019lyapunov} (in that work corresponding to $L\approx9$ with odd-periodic boundary conditions) but this is unsurprising given the localised nature of the chaotic attractor and the very limited range of $L$ for which it exists.

In order to be able to embed the chaotic attractor in a space of dimension 3, it is necessary but not sufficient for the fractal dimension to be less than three. Consider for example the Klein bottle, whose dimension is only two but cannot be embedded within a Euclidean space of dimension less than four. Experiments at larger values of $L$, for example the well-studied\citep{lasagna2018sensitivity} $L\approx39$, and the parameter values studied by \citet{dong2018topological} and \citet{siminos2021manifold} of $L\approx 36$, found attractors whose fractal dimension was apparently less than three, but for which we could only produce a non-injective immersion into a 3D space. An injective embedding is necessary for our topological approach to be valid.

\subsection{Numerical solution}
\label{sec:dns}
We discretize the spatial dimension of the anti-symmetric system with $N_x = 32$ points, turning the function $u$ into a 32-dimensional vector $\boldsymbol{u}$. We observe a satisfactory drop in the amplitudes of the Fourier modes for this discretization, suggesting a well-resolved discretization of the PDE. For comparison, \citet{LanCvitanovic2008PRE} also use 32 modes in the $L = 38.5$ system to find periodic orbits, which exhibits stronger chaos \citep{edson2019lyapunov} and smaller-scale spatial features than our system. To calculate the POD modes in \cref{sec:pod} and train the autoencoder in \cref{sec:autoencoder}, we generate a long direct numerical simulation (DNS) of the system using the ETDRK4 scheme\citep{Kassam2005}, with $t_\mathrm{max} = 20000, dt = 0.1$. Due to the small basin of attraction of the chaotic attractor, we pick a point that we know for certain lies on the attractor as initial condition for the DNS. 

\begin{figure}
    \centering
    \includegraphics[width = \columnwidth]{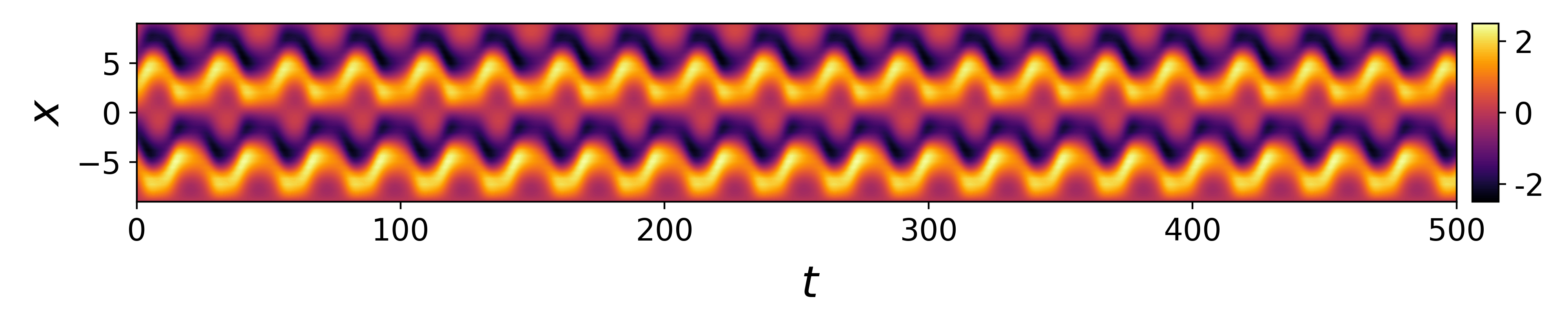}
    \caption{A typical timeseries for the localised chaotic attractor at $L=2\pi/\sqrt{0.1212}$ in the Kuramoto-Sivashinski equation \cref{eq:kse}. To the naked eye the trajectory appears to be periodic, but in fact the small changes in the amplitude of each oscillation are chaotic, as proven by \citet{WILCZAK20208509}.}
    \label{fig:timeseries}
\end{figure}

A section of this timeseries is shown in figure \ref{fig:timeseries}, in the typical manner of plotting space along the vertical axis and using a colour scale to show the value of $u$ as time varies. Unlike the usual projection of trajectories in the Lorenz system, for example, no topology for the dynamics is evident in this figure. Indeed, for this value of $L$ it is not even obvious that the dynamics is chaotic.

\section{State space reduction}
\label{sec:reduction}

\subsection{Proper orthogonal decomposition}
\label{sec:pod}
The most common dimensionality-reduction techniques are linear methods based on eigen-analysis such as dynamic mode decomposition (DMD) \citep{schmid2022dynamic} and principal component analysis (PCA) \citep{Pearson1901, Hotelling1936}, better known as proper orthogonal decomposition (POD) in the fluid dynamics community. POD consists of determining the dominant (uncorrelated) modes that capture most of the system's variance, or `kinetic energy'. Given a timeseries $\{\boldsymbol{u}_i\}_{i = 1}^r$ with $r$ time-steps, where $\boldsymbol{u}_i\in \mathbb{R}^{N_x}$ we can calculate the POD modes by stacking the zero-mean timeseries $\boldsymbol{\tilde{u}}_i = \boldsymbol{u}_i - \boldsymbol{u_\mathrm{mean}}$ in a matrix $\boldsymbol{\tilde{U}}\in\mathbb{R}^{r\times N_x}$  (where the rows are $r$ time-steps), and by considering its covariance matrix. The unbiased estimator $\boldsymbol{C}\in \mathbb{R}^{N_x\times N_x}$ for the covariance matrix is given by 
\begin{equation}
    \boldsymbol{C} = \frac{1}{r-1} \boldsymbol{\tilde{U}}^T\boldsymbol{\tilde{U}}
\end{equation}

The POD modes $\boldsymbol{\phi}_1, ..., \boldsymbol{\phi}_{N_x}$ are the eigenvectors of $\boldsymbol{C}$

\begin{equation}
    \boldsymbol{C}\boldsymbol{\phi}_k = \lambda_k\boldsymbol{\phi}_k
\end{equation}
with corresponding eigenvalues $\lambda_1, ...,\lambda_{N_x}$. The POD modes can then be interpreted as the fluctuations around the mean flow.

\begin{figure}
    \centering
    \includegraphics[width = \columnwidth]{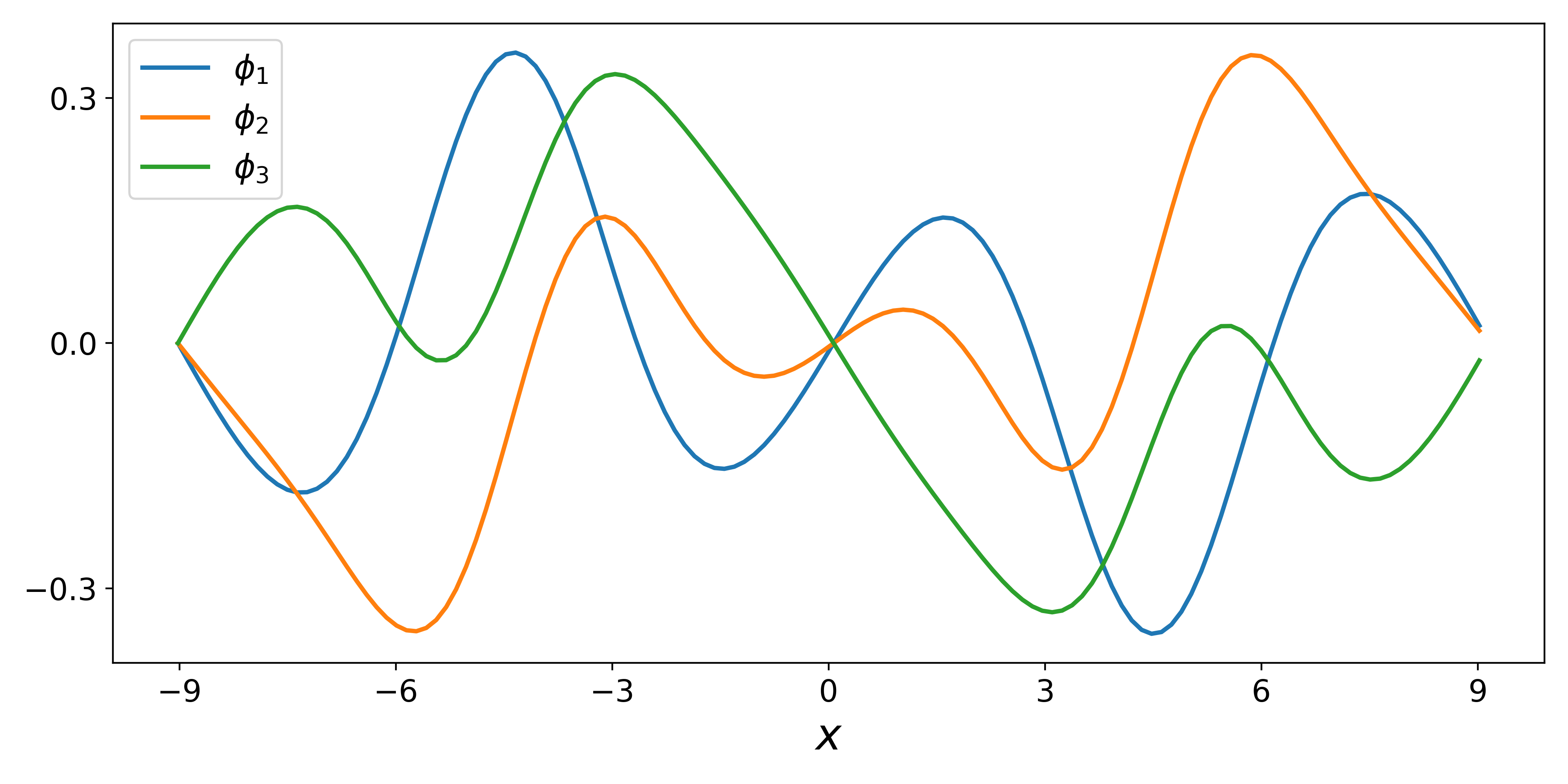}
    \caption{The first three orthonormal POD modes $\phi_1, \phi_2, \phi_3$ of a long time series on the chaotic attractor, with respective eigenvalues $\lambda_1 \approx 19.49, \lambda_2 \approx 1.92$ and $\lambda_3 \approx 1.10$. Since the dynamics are confined to the antisymmetric subspace, all POD modes are odd functions.}
    \label{fig:pod_modes}
\end{figure}

\begin{figure}
    \centering
    \includegraphics[width = \columnwidth]{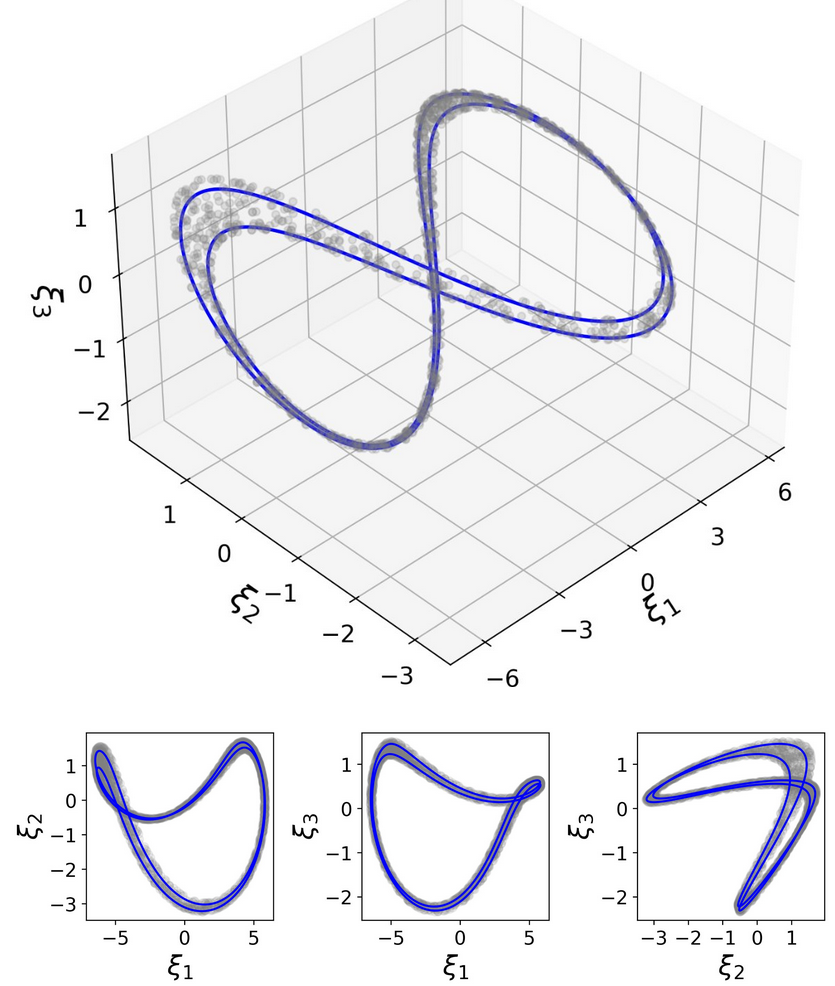}
    \caption{3D plot (top) and 2D projections (bottom) of the chaotic attractor when projected onto the first three POD modes. Formally, we obtain the POD latent coordinates via the inner product $\xi_i = \int_{-L/2}^{L/2} u(x,t) \phi_i(x) dx$. Practically, we obtain them from the dot product $\xi_i(t) = \boldsymbol{u}(\boldsymbol{x},t)\cdot\boldsymbol{\phi}_i$. The grey points are from a timeseries solving the PDE that is projected onto the POD modes. The blue line is a periodic orbit with period $T \approx 51.62$, plotted to emphasise the shape of the attractor.}
    \label{fig:pod_attractor}
\end{figure}

Using a long timeseries from a simulation of the KSE as described in section \ref{sec:dns},
we find the three leading eigenvalues to be $\lambda_1 \approx 19.49, \lambda_2 \approx 1.92$ and $\lambda_3 \approx 1.10$, with the corresponding modes plotted in figure \ref{fig:pod_modes}. The projection of the  attractor defined by the first three POD modes is plotted in figure \ref{fig:pod_attractor} (together with one UPO to emphasise the shape). We obtain the POD latent coordinates $(\xi_1, \xi_2, \xi_3)$ by projecting onto the POD modes. Formally, this is done by evaluating the integral $\xi_i(t) = \int_{-L/2}^{L/2} u(x,t) \phi_i(x) dx$. However, since the spatial dimension is discretized, we obtain the latent POD coordinates from the vector dot product $\xi_i(t) = \boldsymbol{u}(\boldsymbol{x},t)\cdot\boldsymbol{\phi}_i$  Note that the terminology of `latent space' for the expansion into POD modes is chosen in analogy to the alternative autoencoder-based dimensionality reduction approach discussed below. 

The three leading POD modes capture 99.3\% of the variance, which is sufficiently high that we hope the map from the full space to the coefficients of these three modes preserves the topology, and thus allows consistent application of the methods of section \ref{sec:template}.

\subsection{Autoencoder}
\label{sec:autoencoder}

While POD projections capture a great amount of information, they are known to generalise less well to highly nonlinear systems and are outperformed by \textit{autoencoders} \citep{LinotJFM2023}. 
Autoencoders are neural networks that consist of two parameterised functions, namely the \textit{encoder} $\mathcal{E}:\mathbb{R}^{N_x}\rightarrow\mathbb{R}^{N_h}$, where $N_h$ is the dimension of the \textit{latent space}, and the \textit{decoder} $\mathcal{D}:\mathbb{R}^{N_h}\rightarrow\mathbb{R}^{N_x}$. Typically, $N_h\ll N_x$. The parameters are trained so that the autoencoder approximates the identity. That is, for an input vector $\boldsymbol{u}$, we have $\left(\mathcal{D}\circ\mathcal{E}\right)(\boldsymbol{u})\approx\boldsymbol{u}$. Thus, the encoder reduces the dimension of the input, while the decoder does the inverse operation and increases it again. The structure of our autoencoder was chosen so that $\mathcal{E}$ and $\mathcal{D}$ are smooth functions. The goal is to choose $N_h$ as small as possible whilst still preserving the topology of the system. If it is possible to use $N_h=3$ such that $\mathcal{E}$ is an embedding, invertible for points on the attractor, the methods of section \ref{sec:topology} can be applied.

\begin{figure}
    \centering
    \includegraphics[width = 0.9\columnwidth]{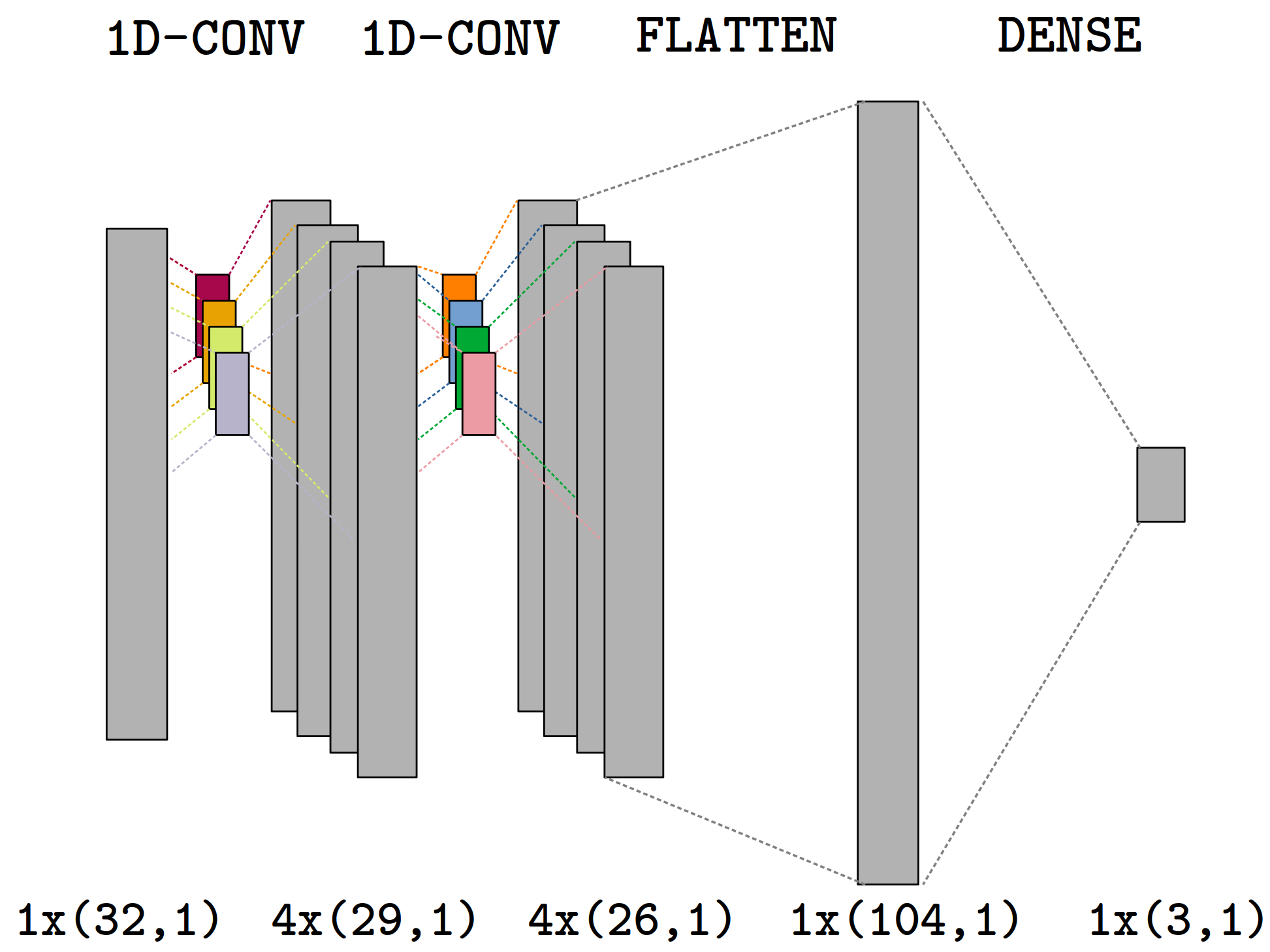}
    \caption{Schematic representation of the encoder $\mathcal{E}$ for $N_x = 32$ and $N_h = 3$: two 1D-convolutional layers are followed by a flatten and a dense layer. The associated decoder $\mathcal{D}$ (not pictured) has the inverse setup: a dense layer, followed by a reshape layer and two 1D transpose convolutional layers.}
    \label{fig:encoder}
\end{figure}

\begin{figure}
    \centering
    \includegraphics[width = \columnwidth]{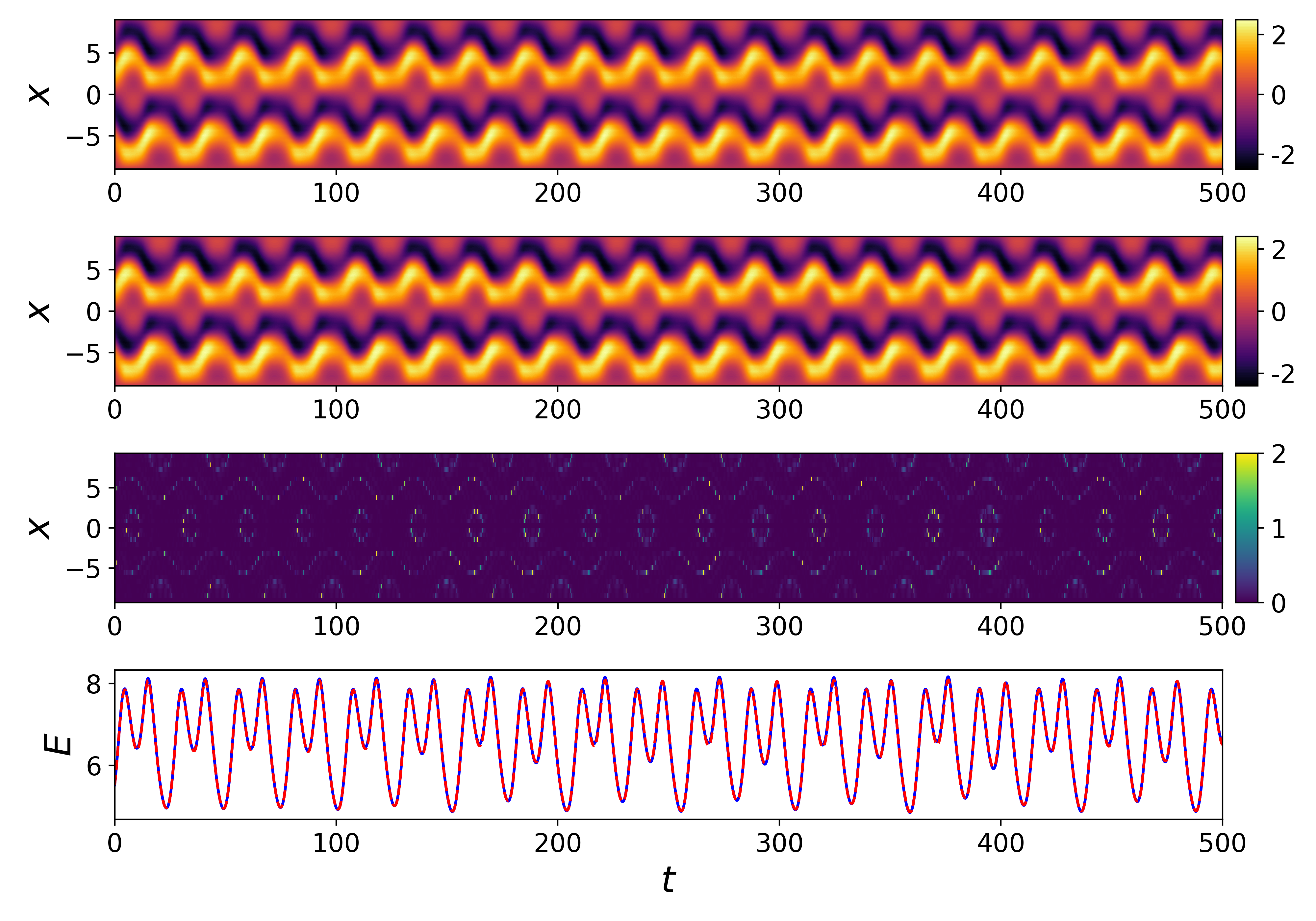}
    \caption{Test performance of the autoencoder applied to a system manifesting low-dimensional chaos. Top: Space-time plot of a DNS that the autoencoder has not seen during training. Second: The autoencoder output of the DNS. Third: The pointwise relative difference between the DNS and the autoencoder output. Yellow dots indicate a relative difference $\geq 2$, which happens when $u(x,t)$ is close to 0. Bottom: Plot of the energy of the DNS (blue solid) and of the autoencoder output (red dashed). At first glance, the signal appears periodic, but comparing the heights of the different peaks makes the chaos visible.}
    \label{fig:test_performance}
\end{figure}

The design of the two neural networks follows those of \citet{beck}. The architecture we pick for the encoder consists of two 1D-convolutional layers, with 4 filters (size 4, stride 1) each, followed by a flatten layer and a dense layer that reduces the dimension to $N_h = 3$ (see figure \ref{fig:encoder} for an illustration of the encoder). Convolutional layers tend to pick up translational invariances and are also sparser than purely dense layers\citep{deeplearningbook}. Since the continuous spatial translation invariance is already discretized to $x \mapsto x + L/2$ by working in the anti-symmetric subspace, the system lacks large-scale invariances, but on small scales there is certainly structure that the network can exploit. The decoder has the inverse setup of the encoder: a dense layer, followed by a reshape layer and two transpose convolutional layers. As activations for each layer, we use the sigmoid function $\sigma(x) = 1/(1 + e^{-x})$.
As mentioned in \ref{sec:dns}, the data used to train the network is a long DNS of the attractor. Before we use the data for training, we re-scale it to the $[0,1]$ interval by applying min-max re-scaling to the zero-mean flow: $\boldsymbol{u}^* = (\boldsymbol{u} - \boldsymbol{u}_\mathrm{mean} - \boldsymbol{u}_\mathrm{min})/(\boldsymbol{u}_\mathrm{max} - \boldsymbol{u}_\mathrm{min})$ (for what follows, we drop the $^*$ for convenience). As loss function, we use the relative error between the input and output rather than the standard mean-squared error, as dividing by the norm of $\boldsymbol{u}$ scales the loss in an interpretable manner. For data points $\{\boldsymbol{u}_i\}_{i = 1}^p$, the loss is:

\begin{equation}
    \mathcal{L} = \frac{1}{p}\sum_{i=1}^p\frac{||\mathcal{D}\circ\mathcal{E}(\boldsymbol{u}_i) - \boldsymbol{u}_i||}{||\boldsymbol{u}_i|| + \epsilon}
\end{equation}
where $\epsilon$ is a small constant to avoid division by $0$. For what follows we continue with a trained autoencoder with a final training loss of $\mathcal{L}=1.41\times 10^{-4}$ and final test loss of $\mathcal{L}=1.45\times 10^{-4}$, indicating good out-of-sample generalization. The test performance of the autoencoder is illustrated in figure \ref{fig:test_performance}.

\begin{figure}
    \centering
    \includegraphics[width = \columnwidth]{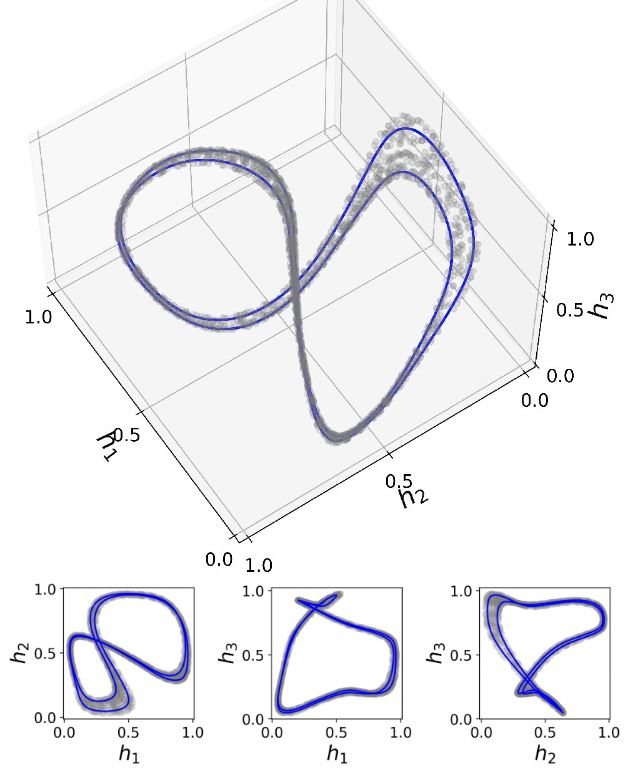}
    \caption{3D plot (top) and 2D projections (bottom) of the chaotic attractor in the autoencoder's latent space. The latent coordinates $(h_1, h_2, h_3)$ are obtained by applying the encoder $\mathcal{E}$ to a data point. The grey points are from a timeseries solving the PDE. The blue line is a periodic orbit with period $T \approx 51.62$, plotted to emphasise the shape of the attractor.}
    \label{fig:ae_attractor}
\end{figure}

Since we define the latent dimension to be $N_h = 3$, this gives us the desired 3-dimensional representation of the KSE, in which we would like to test the methods described in section \ref{sec:topology}. Taking datapoints from a long physical timeseries and applying the encoder, we obtain a 3D latent timeseries. Figure \ref{fig:ae_attractor} shows a 3D plot of the image of the attractor in latent space, illustrating its band-like shape, and 2D projections. 

\subsection{Poincaré section and return maps}
\label{sec:poincare}

In the previous section we described how proper orthogonal decomposition and autoencoders allow us to obtain 3-dimensional approximate coordinates for the KSE, denoted by $(\xi_1, \xi_2, \xi_3)$ and $(h_1, h_2, h_3)$ respectively. In this section, for each of the state space reduction techniques, we will define a Poincaré section and a return map. These will help us with identifying UPOs and allow us to obtain a set of names for them. 

\subsubsection{POD return map}
\begin{figure}
    \centering
    \includegraphics[width = 0.8\columnwidth]{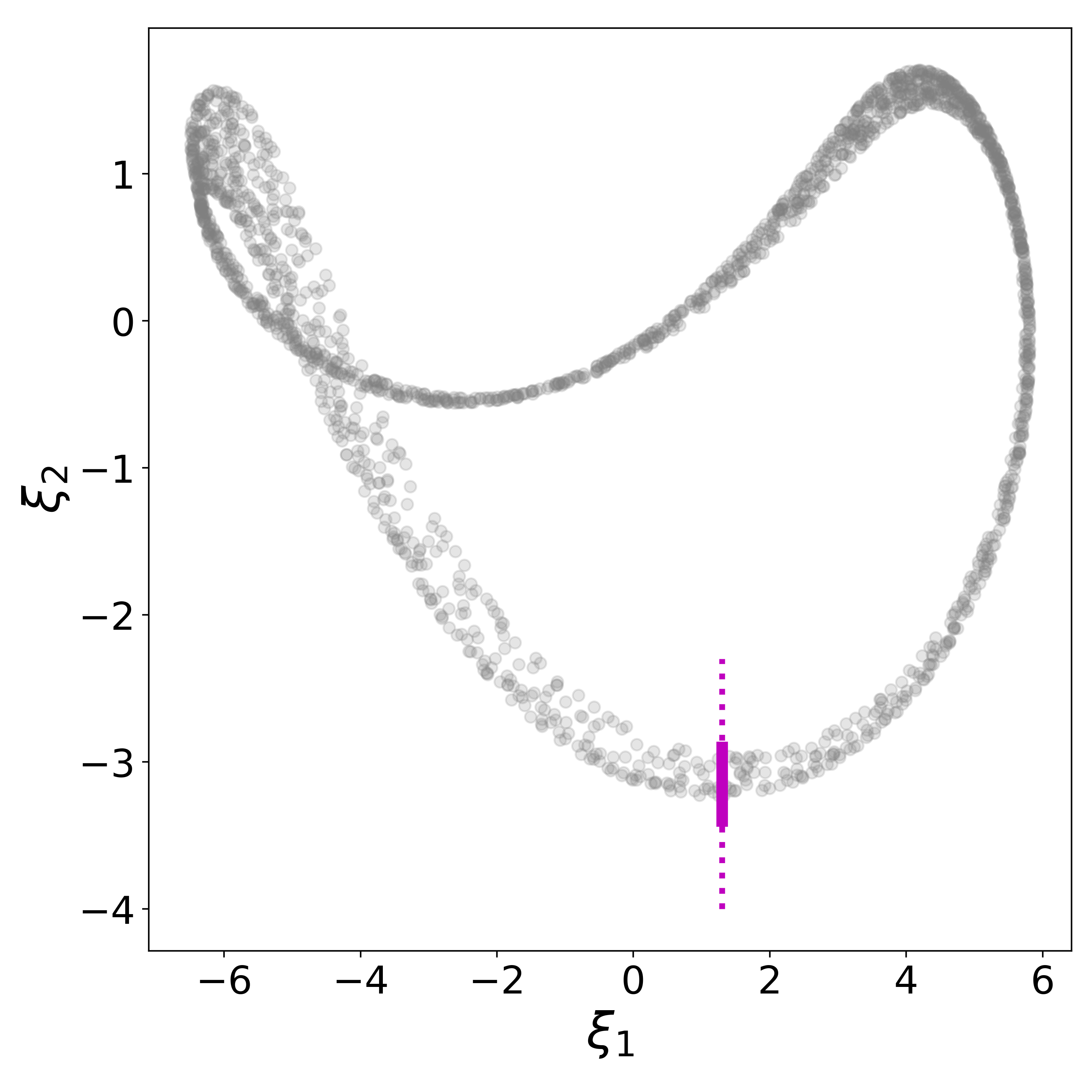}
    \caption{$(\xi_1, \xi_2)$ projection of the chaotic attractor in the latent space defined by the POD coordinates. The grey points are from a physical timeseries that is put through the encoder. The magenta line illustrates the Poincaré section $\xi_1 = 1.3$ and $\partial_t \xi_1 < 0$, denoted $\mathcal{P}_{\mathrm{POD}}$.}
    \label{fig:pod_poincare_2d}
\end{figure}

\begin{figure}
    \centering
    \includegraphics[width = \columnwidth]{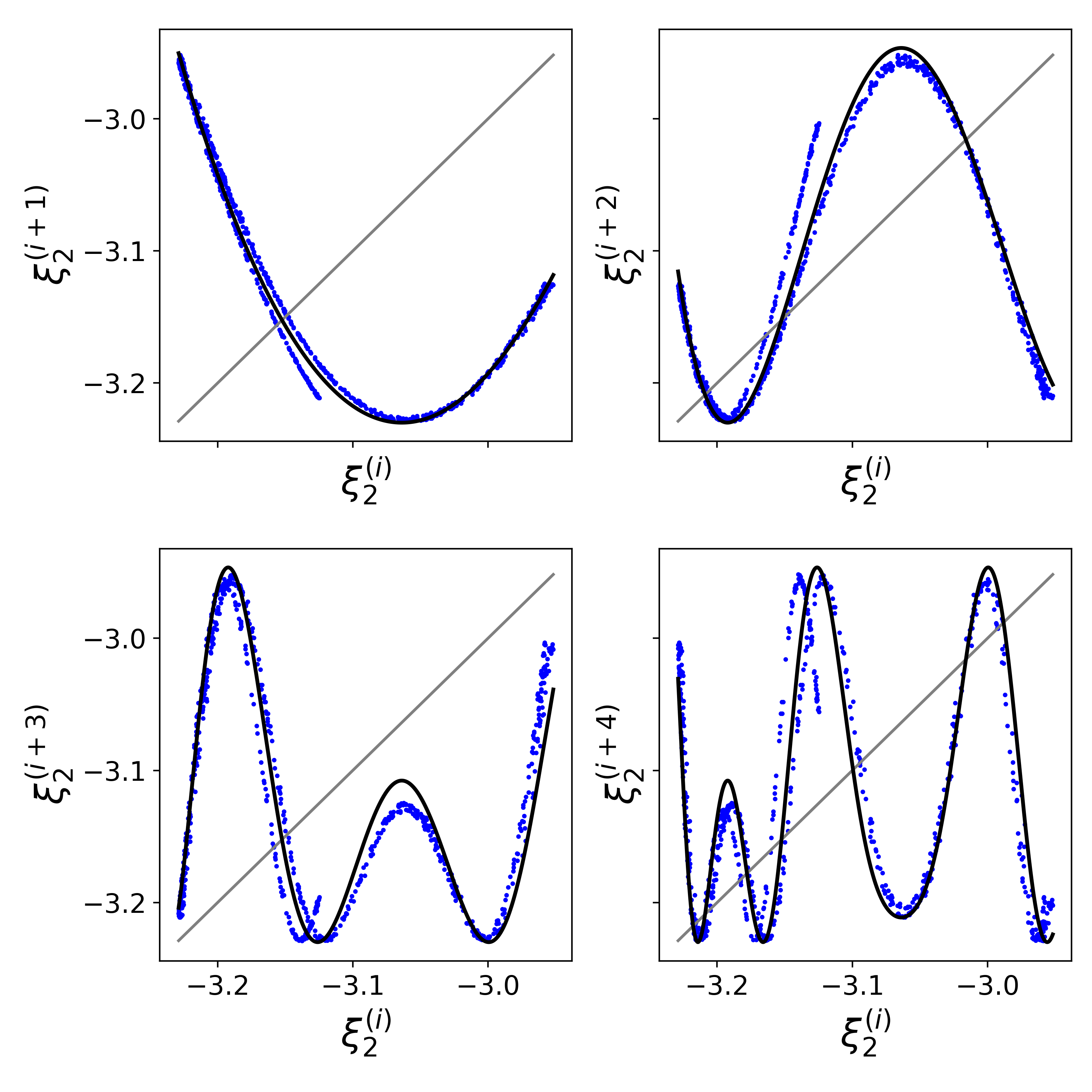}
    \caption{Blue: Scatter plots of subsequent passages through the Poincaré section $\mathcal{P}_{\mathrm{POD}}$ indicating the return map (top left) and its second (top right), third (bottom left) and fourth (bottom right) iterate. Black: third order polynomial fitting $f_1:\xi_2^{(i)}\mapsto \xi_2^{(i+1)}$ of the first return map, as seen in the top left tile. The black curves in the other tiles are obtained by applying $f_1$ multiple times, e.g. $f_2(\xi_2^{(i)}):=f_1(f_1(\xi_2^{(i)}))$ is plotted in the top right tile. Intersections of the fitted (iterated) return map(s) with the identity suggest the existence of periodic orbits crossing the Poincaré section $p=1,2,4$ times, but there is no evidence for a periodic orbit with $p=3$ crossings. }
    \label{fig:pod_return_map}
\end{figure}

\begin{figure}
    \centering
    \includegraphics[width = \columnwidth]{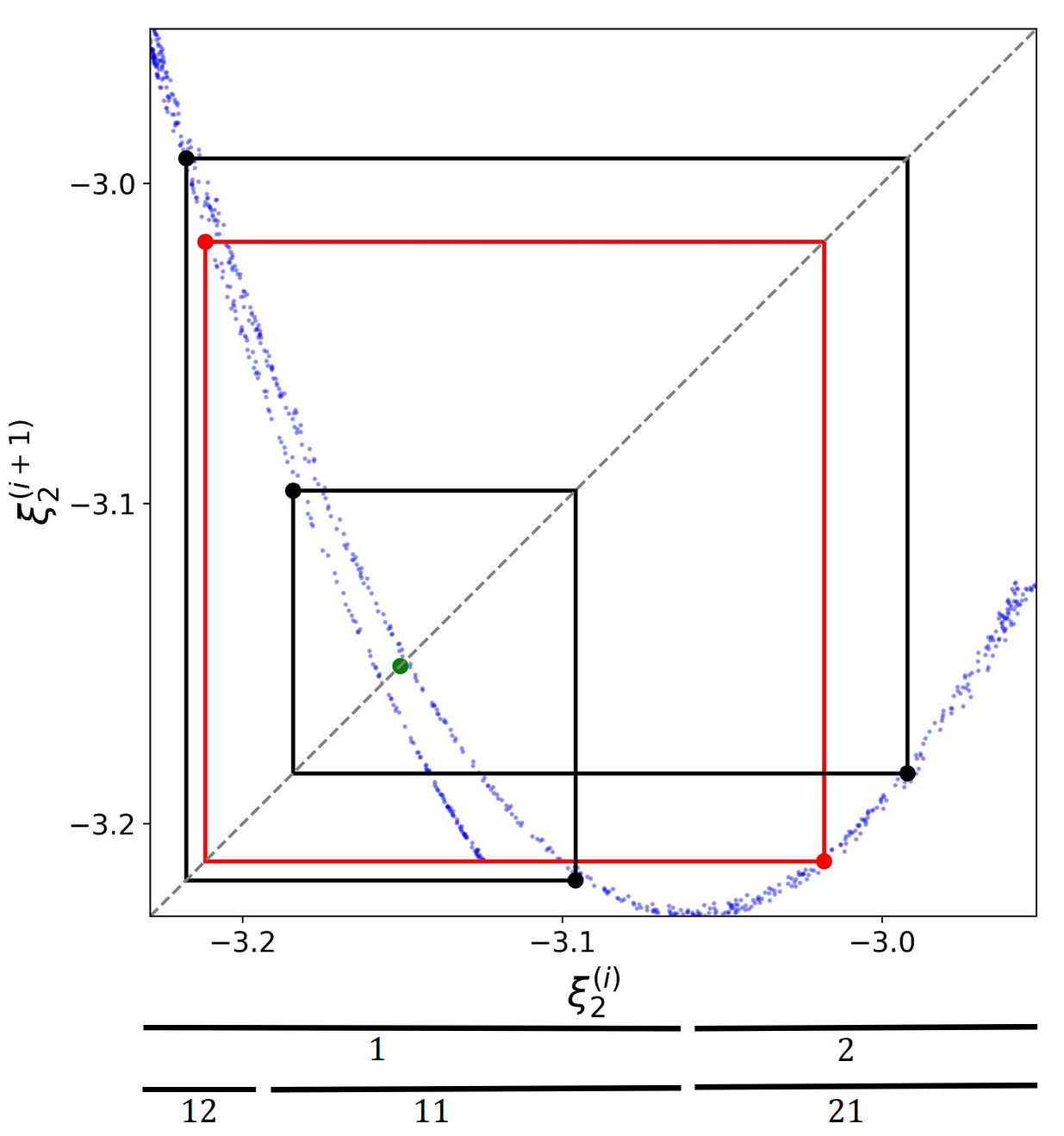}
    \caption{First return map for the Poincaré section $\mathcal{P}_{\mathrm{POD}}$. The blue dots show data from a chaotic timeseries, whereas the green dot, red line and black line show the periodic orbits $1_1$, $2_1$ and $4_1$ respectively (see \cref{tab:UPOs}). Below the figure, the intervals represent the first and second level within the state space partition derived from the simplified return map shown in figure \ref{fig:return_map}. These intervals are used to construct symbolic sequences for the periodic orbits.}
    \label{fig:pod_first_return_map}
\end{figure}

In the POD latent coordinates $(\xi_1, \xi_2, \xi_3)$, we define a Poincaré section $\mathcal{P}_{\mathrm{POD}}$ given by the plane $\xi_1 = 1.3$ and flow direction $\partial_t \xi_1 < 0$. The value $\xi_1 = 1.3$ is chosen so that the flow is approximately orthogonal to the Poincaré section. This defines a unique crossing section along the band (see figure \ref{fig:pod_poincare_2d}), and gives us a sequence $\{\xi_2^{(i)}\}$ of the $\xi_2$ coordinate of the crossings through $\mathcal{P}_{\mathrm{POD}}$. The return map $f_1$ is defined by $f_1: \xi_2^{(i)}\mapsto \xi_2^{(i + 1)}$. A plot of such subsequent crossings is shown in the top left panel of figure \ref{fig:pod_return_map}. The points approximately fall on a smooth unimodal curve, which we fit with a polynomial of order three. This simplified first return map appears to only have one intersection with the identity $\xi_2^{(i)} = \xi_2^{(i + 1)}$, suggesting that there is only one periodic orbit with $p=1$, where $p$ is the number of intersections of an orbit with $\mathcal{P}_{\mathrm{POD}}$. By looking at further iterations/compositions of the return map (see other panels of figure \ref{fig:pod_return_map}), we also obtain an indication that there are 1, 0 and 1 periodic orbits with $p = 2,3,4$ respectively (accounting for crossings that correspond to periodic orbits with smaller $p$ - e.g. the orbit with $p=1$ will also account for one intersection in all subsequent graphs). Finally, we split the band domain $I = [-3.229, -2.952]$ at the minimum value of the first return map, giving approximately $I_1 = [-3.229,-3.055]$ and $I_2 = [-3.055, -2.952]$. From this, we obtain an indication of the qualitative dynamics: points starting in $I_1$ are mapped to the full interval $I$, while points starting in $I_2$ are mapped to a subset of $I_1$. This is shown in figure \ref{fig:pod_first_return_map}.

\subsubsection{Autoencoder return map}
\begin{figure}
    \centering
    \includegraphics[width = 0.8\columnwidth]{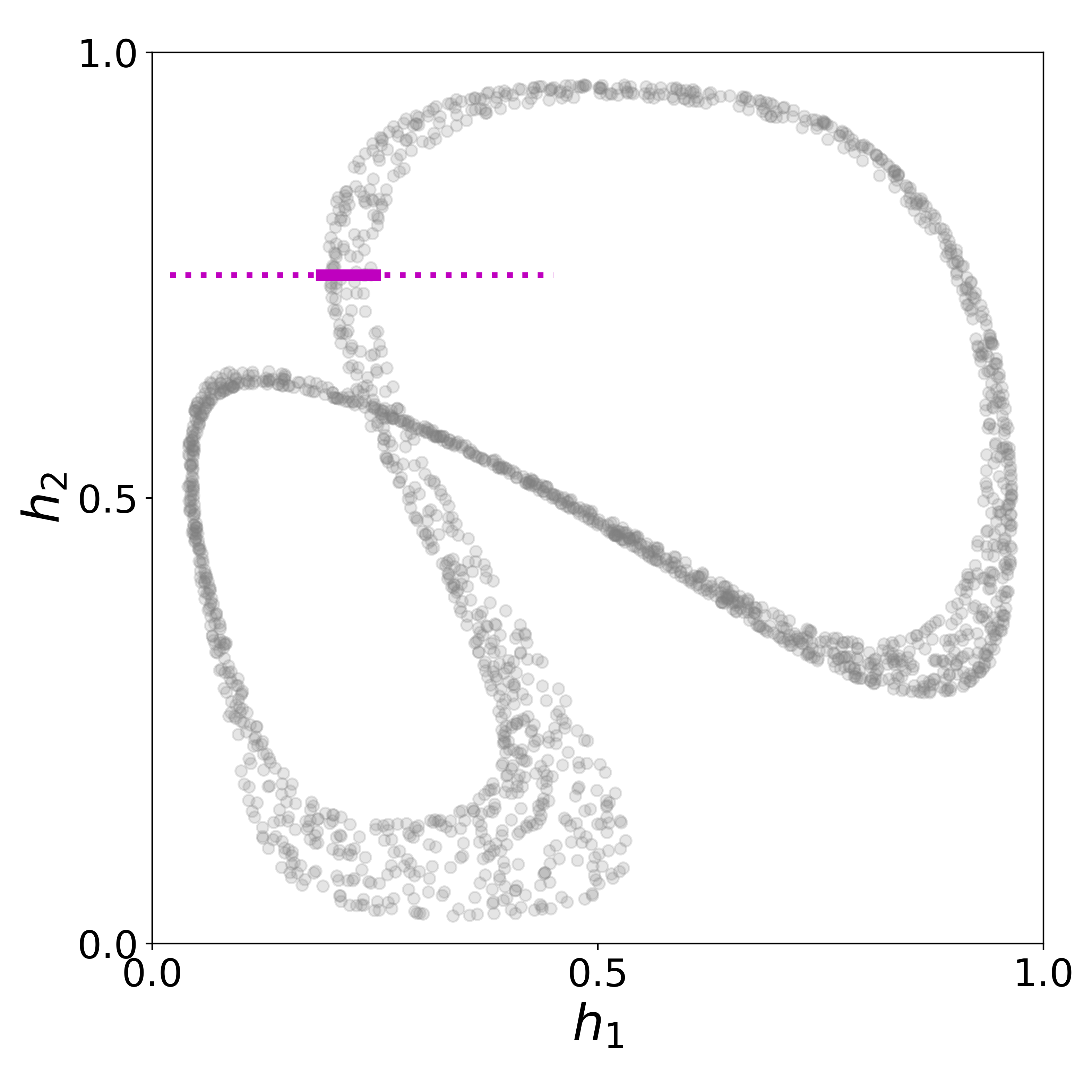}
    \caption{$(h_1, h_2)$ image of the chaotic attractor in the latent space of the autoencoder. The grey points are from a physical timeseries that is put through the encoder. The magenta line illustrates the Poincaré section $h_2 = 3 / 4$ and $\partial_t h_2 >0$, denoted $\mathcal{P}_{\mathrm{AE}}$.}
    \label{fig:poincare_2d}
\end{figure}

\begin{figure}
    \centering
    \includegraphics[width = \columnwidth]{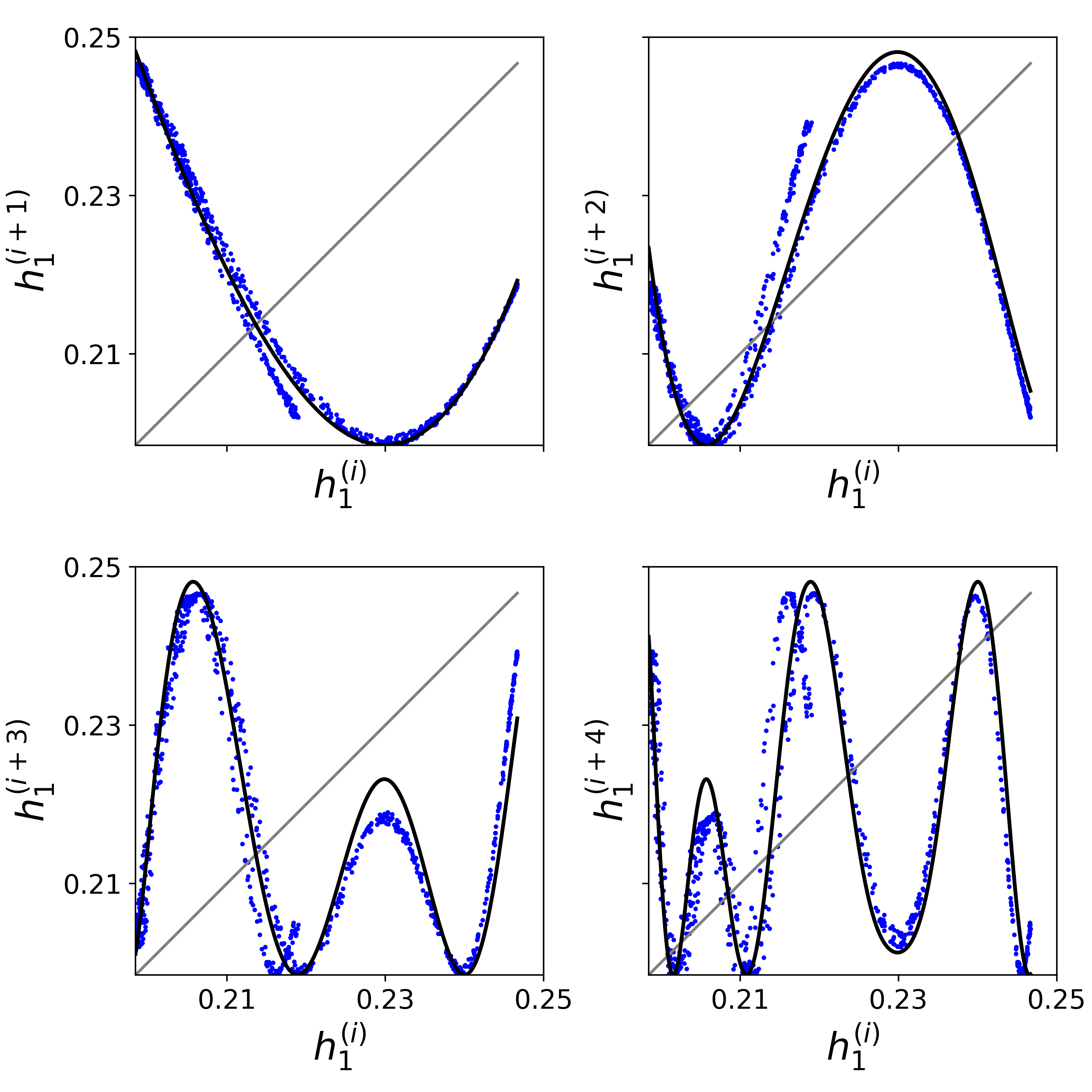}
    \caption{Blue: Scatter plots of subsequent passages through the Poincaré section $\mathcal{P}_{\mathrm{AE}}$ indicating the return map (top left) and its second (top right), third (bottom left) and fourth (bottom right) iterate. Black: third order polynomial fitting $g_1:h_1^{(i)}\mapsto h_1^{(i+1)}$ of the first return map, as seen in the top left tile. The black curves in the other tiles are obtained by applying $g_1$ multiple times, e.g. $g_2(h_1^{(i)}):=g_1(g_1(h_1^{(i)}))$ is plotted in the top right tile. Intersections of the fitted (iterated) return map(s) with the identity suggest the existence of periodic orbits crossing the Poincaré section $p=1,2,4$ times, but there is no evidence for a periodic orbit with $p=3$ crossings. }
    \label{fig:return_map}
\end{figure}

\begin{figure}
    \centering
    \includegraphics[width = \columnwidth]{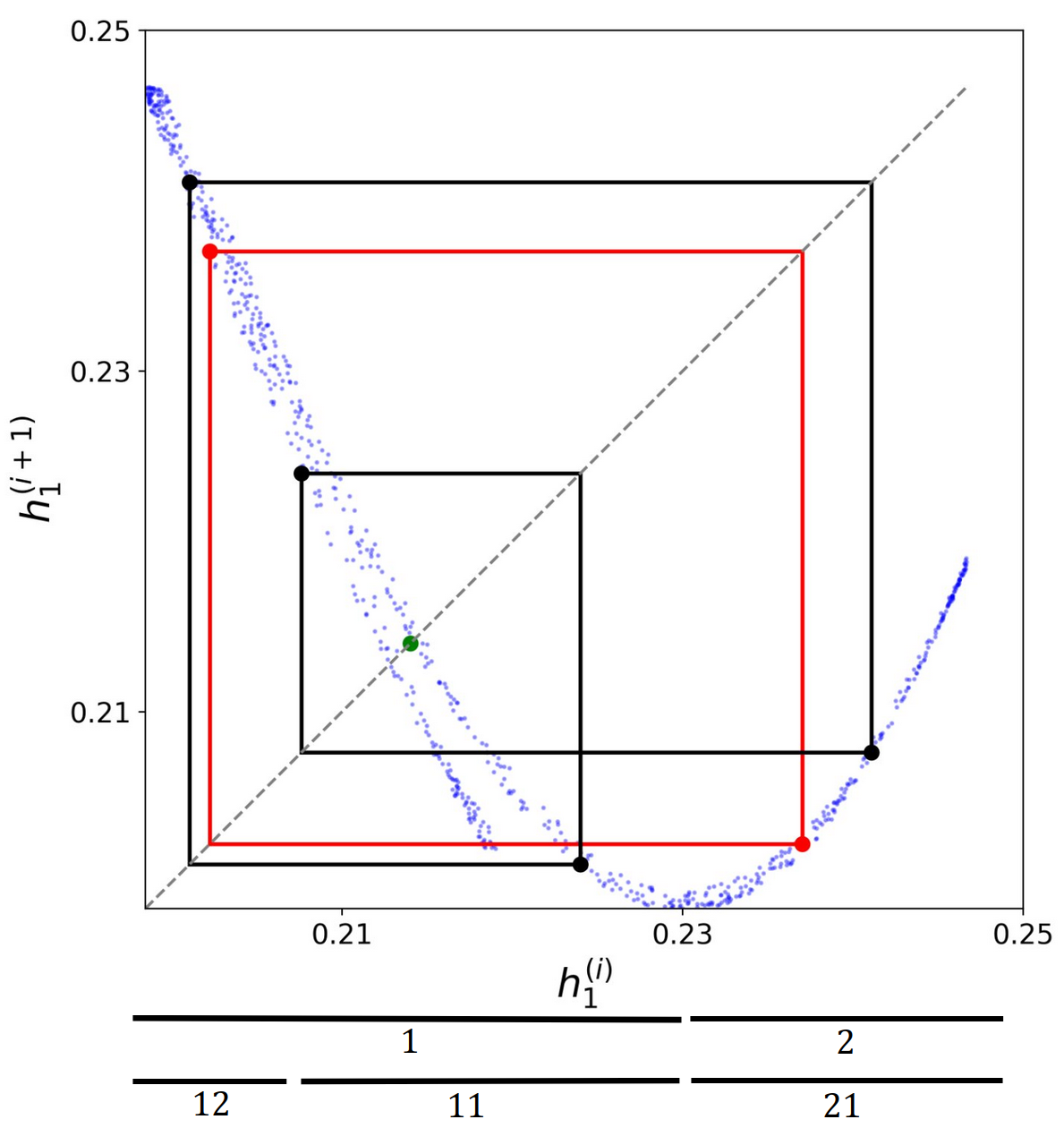}
    \caption{First return map for the Poincaré section $\mathcal{P}_{\mathrm{AE}}$. The blue dots show data from a chaotic timeseries, whereas the green dot, red line and black line show the periodic orbits $1_1$, $2_1$ and $4_1$ respectively (see \cref{tab:UPOs}). Below the figure, the intervals represent the first and second level within the state space partition derived from the simplified return map shown in figure \ref{fig:return_map}. These intervals are used to construct symbolic sequences for the periodic orbits.}
    \label{fig:first_return_map}
\end{figure}

We proceed in the same way with the autoencoder's latent coordinates $(h_1, h_2, h_3)$ . We define a Poincaré section $\mathcal{P}_{\mathrm{AE}}$ given by the plane $h_2 = 3 / 4$ and flow direction $\partial_t h_2 >0$, so that the flow is approximately orthogonal to it. Figure \ref{fig:poincare_2d} shows that this defines a unique crossing section along the band, giving us a sequence $\{h_1^{(i)}\}$. The return map $g_1$ is defined by $g_1: h_1^{(i)}\mapsto h_1^{(i + 1)}$ (see top left panel of figure \ref{fig:return_map}). We again fit the resulting smooth curve, with a polynomial of order three. We make the same observations as for the POD return map: $g_1$ seems to have one intersection with the identity $h_1^{(i)} = h_1^{(i + 1)}$, implying one periodic orbit with $p=1$. Further iterations/compositions of the return map (see figure \ref{fig:return_map}) indicate that there are 1, 0 and 1 periodic orbits with $p = 2,3,4$ respectively. We again split the band domain $J = [0.198, 0.247]$ at the minimum value of $g_1$, giving approximately $J_1 = [0.198, 0.23]$ and $J_2 = [0.23, 0.247]$. We observe similar qualitative dynamics to the POD return map: points starting in $J_1$ are mapped to $J$, while points starting in $J_2$ are mapped to a subset of $J_1$ (see figure \ref{fig:first_return_map}).

\subsection{Finding periodic orbits}
\label{sec:finding_upos}
We find periodic orbits (such as the one plotted in figures \ref{fig:pod_attractor} and \ref{fig:ae_attractor}) by generating guesses using the methodology from \citet{beck} and converging them using the algorithms from \citet{Azimi2022}. This is similar to the variational method used by \citet{dong2018topological} to find UPOs and symbolic dynamics at a higher $L$ value in the KSE. We obtain the guesses by defining arbitrary closed curves (loops) within the latent space, and decoding them to the physical space. This gives us a timeseries that is already time-periodic and lies close on the chaotic attractor, however does not necessarily satisfy the PDE. The convergence algorithm then deforms the loop in an attempt to turn it into a solution, by minimizing a cost function $J$ that quantifies the misalignment between the tangent vectors and the flow vectors at each point of the loop. Thus, a root of the cost function corresponds to a closed curve in the state space that satisfies the flow equations everywhere, and is hence a periodic orbit.

Following this methodology, we find many periodic orbits of the system, not all of which necessarily lie on the chaotic attractor that we are considering in this paper. To characterize the attractor, we need to ensure that only those periodic orbits that lie on the attractor are considered in the analysis. Since the latent space is 3-dimensional, we can plot the attractor with the found periodic orbits. Visual inspection allows us to discard those orbits that clearly do not lie on the attractor. In our analysis, we thus only consider the periodic orbits that appear to lie on the chaotic attractor in the latent space (of which we find 15 with $p < 12$). A list of periods $T$ and intersections $p$ with $\mathcal{P}_{\mathrm{POD}}$ and $\mathcal{P}_{\mathrm{AE}}$ (as defined in section \ref{sec:poincare}) is given in table \ref{tab:UPOs}.

Comparing the extracted list of 15 periodic orbits with the return map and its iterates in figure \ref{fig:return_map} suggests that we have successfully identified all relevant periodic orbits with $p < 7$.  
For example, we initially find periodic orbits with $p = 3$, but upon visual inspection, we discarded all of those as not lying on the attractor. 
This agrees with our return maps. Since the only intersection of the third iterate with the identity (in the bottom left panel of figure \ref{fig:return_map}) corresponds to the $p = 1$ orbit, the iterated return map indeed suggests that no orbit with $p=3$ on the attractor exists. This is consistent with the symbolic dynamics derived from the approximate return map as shown in figure \ref{fig:first_return_map}, since period 3 symbolic sequences are not admissible \citep{hao1998applied}. Recall that the existence of a period 3 orbit implies chaos, but the converse is not true. Likewise, all identified orbits exactly correspond to those expected from the return maps. For $p \geq 7$, it is difficult to be certain about the number of periodic orbits by purely looking at the return maps, and so we do not claim that we have found all of them with certainty. All orbits that appear to lie on the attractor were included in our analysis, and none were discarded purely on the basis of a return map argument.
For the $15$ orbits, we arbitrarily chose names based on $p$ and a running index, as indicated in table \ref{tab:UPOs}.

\begin{table}
    \begin{center}
        \begin{tabular}{ |c|c|c|c| } 
            \hline
            $T$ & $p$ & Name & Symbolic sequence \\
            \hline
            25.706 & 1 & $1_1$ & $(1)$ \\
            51.617 & 2 & $2_1$ & $(21)$ \\
            103.190 & 4 & $4_1$ & $(2111)$ \\
            128.997 & 5 & $5_1$ & $(21111)$ \\
            129.036 & 5 & $5_2$ & $(21121) $ \\
            154.731 & 6 & $6_1$ & $(211111)$\\
            154.789 & 6 & $6_2$ & $(211121)$ \\
            180.506 & 7 & $7_1$ & $(2111111)$ \\
            180.558 & 7 & $7_2$ & $(2111121)$ \\
            206.258 & 8 & $8_1$ & $(21111111)$ \\
            206.315 & 8 & $8_2$ & $(21111121)$ \\
            232.192 & 9 & $9_1$ & $(211112121)$ \\
            257.836 & 10 & $10_1$ & $(2111111121)$\\
            283.597 & 11 & $11_1$ & $(21111111121)$ \\
            283.801 & 11 & $11_2$ & $(21111212121)$ \\
            
            \hline
        \end{tabular}
    \end{center}
\caption{Period $T$ and number of Poincaré intersections $p$ with the Poincaré sections $\mathcal{P}_{\mathrm{POD}}$ and $\mathcal{P}_{\mathrm{AE}}$ for each of the 15 periodic orbits found for the chaotic attractor of interest. The names are chosen based on $p$ and a running index in the order of increasing period $T$. Two identical set of symbolic sequences is obtained from the first return maps for their respective Poincaré sections $\mathcal{P}_{\mathrm{POD}}$ and $\mathcal{P}_{\mathrm{AE}}$.}
\label{tab:UPOs}
\end{table}

\section{Topological Analysis }

The arrangements of orbits embedded within strange attractors of $3$D dynamical systems provides topological invariants -- properties of the system which are same, regardless of the particular choice of coordinates. \citet{Birman1983KnottedPO} showed that, for a hyperbolic system in three dimensions, a strange attractor projects onto a simpler object, called a \textit{knot holder} or \textit{template} \cite{birman1983knotted} which is obtained by identifying (in the mathematical sense) points which lead to the same trajectories in the long-time limit via the relation: \[x\sim y \Leftrightarrow \lim_{t\longrightarrow \infty} \left\Vert x(t) - y(t) \right\Vert = 0. \] With one negative Lyapunov exponent indicating one contracting direction, this identification procedure reduces the dimension by one: the template is a two-dimensional branched manifold, i.e. a two-dimensional manifold everywhere except at those locations where two-dimensional branches separate (a stretching mechanism) or join (a squeezing mechanism)\citep{ReviewGilmore}. Considering such a 2D structure was inspired by the observation that strange attractors in 3D systems often appear to be very thin: they are `almost' two-dimensional. In a real dynamical system they cannot be perfectly two-dimensional, since separation of two branches would violate uniqueness of trajectories. However, the template encodes key dynamical properties of the attractor. 

The Birman-Wiliam theorem also states that periodic orbits of the strange attractor are in one-to-one correspondence with the periodic orbits on the template, with at most two extraneous orbits\citep{birman1983knotted, Ghrist1997}. Moreover, the template has the property that all periodic orbits can be projected onto it without altering any of their topological invariants, crucially including their (self-)linking numbers. Certain preconditions of the Birman-Wiliam theorem can be relaxed; see Chapter 13 in \citet{TopChaos} for more details. Given a higher dimensional flow whose Lyapunov dimension is less than 3, one can first project the flow along its stable directions to an intermediate manifold where we  can apply the Birman-Wiliam theorem. 
In the following, we attempt to construct the template of the chaotic attractor from the identified set of periodic orbits and thereby identify the structure of the spectrum of periodic orbits on the attractor of the PDE. 

\subsection{Method}
\label{sec:topology}
A branched manifold associated with a template of $n$ branches can be given an algebraic description consisting of a small number of integers: torsion and layering numbers\citep{TopChaos}. The torsion terms are encapsulated in a $n\times n$ symmetric matrix of integers describing how the branches are knotted together. The entry $t_{i,j}$ is the signed number of crossings between branch $i$ and branch $j$, and $t_{i,i}$ is the self-torsion of branch $i$, i.e. signed number of crossings between the two edges of the branch. In addition, the layering of the branches can be encapsulated in a vector of integers, consisting of entries $l_{ij}$ for $i<j$. The entry $l_{i,j}$ for $i<j$, is $1$ if branch $j$ is closer to the reader than branch $i$, and $-1$ otherwise. While this algebraic description is not topologically invariant (it depends on the choice of projection of the template in 2D), it determines the arrangement of the periodic orbits, which is the topological invariant of interest in this work.

These integers describing the template can themselves be determined by computing one of two topological invariants of the periodic orbits: their (self-)linking numbers or relative rotation rates.

The linking number $Lk(A,B)$ of two knotted curves $A$ and $B$ (embedded in three-dimensional Euclidean space) is an integer quantifying the number of times one curve winds around the other. To compute the linking number of two curves, we label each crossing as positive or negative, then the total number of positive crossings minus the total number of negative crossings is equal to twice the linking number. That is,
\begin{align}\label{lknumber}
    Lk(A,B)= \frac{1}{2} \sum_{i}\sigma_i(A,B),
\end{align}
where the signed crossing $\sigma_i(A,B)$ equals $+1$ if the $i$-th crossing between $A$ and $B$ is right-handed and $-1$ if it is left-handed. The linking number is a topological invariant, i.e. it is preserved under continuous deformations. The linking number between closed curves can also be obtained by evaluating the Gauss linking integral\citep{Ricca2011GAUSSLN} \begin{align*}\label{Top_lk}
    Lk(A,B)= \frac{1}{4 \pi} \int_{A}^{} \int_{B}^{}  \frac{\mathbf{r}_1-\mathbf{r}_2}{\lvert \mathbf{r}_1 - \mathbf{r}_2\rvert^3} \cdot \, \mathrm{d}\mathbf{r}_1 \times \mathrm{d}\mathbf{r}_2,
\end{align*}
but this is more complicated to evaluate accurately.

A template provides a way of enumerating the periodic orbits by assigning a unique symbol to each branch, yielding a descriptive name based on the symbolic sequence that the orbit traverses the branches\citep{Plumecoq1, TopChaos}.
Let $A=(a_1,\dots,a_{p_A})$ and $B=(b_1,\dots,b_{p_B})$ be two periodic orbits written with their symbolic sequences. Here, $p_A$ and $p_B$ denote the number of intersection points of the respective orbits with a Poincaré section. The relative rotation rates $R_{i,j}(A,B)$ describe how much, on average, two orbits $A$ and $B$ rotate around one another when starting from initial conditions $a_i$ and $b_j$\citep{RRRSolari}.
\begin{equation}\label{Lk_RRR}
   R_{i,j}(A,B) =  \frac{1}{p_A p_B} \sum_{k=1}^{p_A p_B} \frac{1}{2 }\sigma(A_{i+k},B_{j+k}).
\end{equation}
Here $A_m$ is the segment of $A$ between $a_m$ and $a_{m+1}$ in the Poincaré section and $\sigma(A_{i+k},B_{j+k})$ is the sum of the signed crossings between the segments $A_{i+k}$ and $B_{j+k}$ according to the rule described above. Then, from (\ref{lknumber}) and (\ref{Lk_RRR}) the linking number can be recovered from the relative rotation rates (see Appendix in \citet{RRRSolari}),
\begin{equation*}
   Lk(A,B) =  \sum_{\substack{i=1,\dots,p_A\\j=1,\dots,p_B}}R_{i,j}(A,B).
\end{equation*} 

In this way, the linking numbers can be expressed as polynomials in the torsion and layering terms, with coefficients depending on the symbolic sequence assigned to each orbit (see Appendix A.2.5 in \citet{TopChaos}),
\begin{equation}\label{Formula}
   Lk(a,b) = P_t(t_{i,j}) + P_l(t_{i,j}, l_{i,j}).
\end{equation}

In summary, a template can be determined from an input set of periodic orbits via the following algorithm\citep{TopChaos}:
\begin{itemize}
    \item Compute the linking numbers of the input set of orbits.
    \item If possible, determine the number of branches in the template from the return map. The number of branches is given by the stretching that occurs during one period. It equals the number of monotonic branches of the first return map\citep{TopChaos}.
    \item Iterate over the possible symbolic sequences for the orbits (and/or the the number of branches) and solve the system of equations (\ref{Formula}). If there is a unique valid solution one can include additional orbits to check the validity of a candidate. If, on the other hand, multiple valid names and template candidates are found, the problem is undetermined and more orbits must be included.
\end{itemize} 

Hence the algorithm returns valid set(s) of possible names together with a template. The procedure is described precisely in Appendix A in \citet{TopChaos}. An overview of the topological analysis program is given in \citet{LefrancOVerview}. The previous algorithm does not immediately extend to flows in dimensions higher than three. However, in the considered case of the KSE, it appears that the attractor can be embedded in $\mathbb{R}^3$ where the computations of linking numbers can be carried out.

\subsection{A template for the KSE}
\label{sec:template}

Strange attractors can be classified by the topological organization of periodic orbits\cite{Letellier_2022}. If the dynamics can be embedded in three dimension then the periodic orbits form knots and links and their topological organization is described by their linking numbers. Then, by the Birman-Wiliam theorem, one can project the flow onto a template while preserving the organization of the periodic orbits. A template not only supports all the periodic orbits and describes their topological organization but also offers a natural  way for enumerating these periodic orbits. For $L\approx18.05$, the Lyapunov dimension associated to the KSE system is lower than three. We use two distinct tools: proper orthogonal decomposition and an autoencoder neural network, described earlier, to embed the dynamic in 3D latent spaces. Then we compare the templates resulting from each method.

\begin{table}
\label{tab:linking}
\begin{center}
\begin{tabular}{ |c |c|c|c|c|c|c|c|c|c|c|c|c|c|c|c| } 
\hline
& $1_1$ & $2_1$ & $4_1$ & $5_1$ & $5_2$ & $6_1$ & $6_2$ & $7_1$ & $7_2$ & $8_1$ & $8_2$ & $9_1$ & $10_1$ & $11_1$ & $11_2$  \\
\hline
 $1_1$ &  & -1 & -2 & -3 & -3 & -3 & -3 & -4 & -4 & -4 & -4 & -5 & -5 & -6 & -6\\
 $2_1$ &  &  & -5 & -6 & -6 & -7 & -7 & -8 & -8 & -9 & -9 & -11 & -11 & -12 & -13 \\
 $4_1$ &  &  &  & -12 & -12 & -14 & -14 & -16 & -16 & -18 & -18 & -21 & -22 & -24 & -26\\
 $5_1$ &  &  &  &  & -15 & -18 & -18 & -21 & -21 & -24 & -24 & -27 & -30 & -33 & -33 \\
 $5_2$ &  &  &  &  &  & -18 & -18 & -21 & -21 & -24 & -24 & -27 & -30 & -33 & -33 \\
 $6_1$ &  &  &  &  &  &  & -21 & -24 & -24 & -27 & -27 & -31 & -33 & -36 & -38 \\
 $6_2$ &  &  &  &  &  &  &  & -24 & -24 & -27 & -27 & -32 & -33 & -36 & -39 \\
 $7_1$ &  &  &  &  &  &  &  &  & -28 & -32 & -32 & -36 & -40 & -44 & -44 \\
 $7_2$ &  &  &  &  &  &  &  &  &  & -32 & -32 & -37 & -40 & -44 & -45 \\
 $8_1$ &  &  &  &  &  &  &  &  &  &  & -36 & -41 & -44 & -48 & -50 \\
 $8_2$ &  &  &  &  &  &  &  &  &  &  &  & -42 & -44 & -48 & -51 \\
 $9_1$ &  &  &  &  &  &  &  &  &  &  &  &  & -52 & -57 & -59 \\
 $10_1$ &  &  &  &  &  &  &  &  &  &  &  &  &  & -60 & -63\\
 $11_1$ &  &  &  &  &  &  &  &  &  &  &  &  &  &  & -69\\
 $11_2$ &  &  &  &  &  &  &  &  &  &  &  &  &  &  & \\
\hline
\end{tabular}
\end{center}
\caption{Linking numbers of periodic orbits projected into three dimensions using POD. Using the autoencoder instead, the same linking numbers were found but with positive sign.}
\label{lknbr}
\end{table}

\subsubsection{POD template}
\begin{figure}
    \centering
    \includegraphics[width=0.55\linewidth]{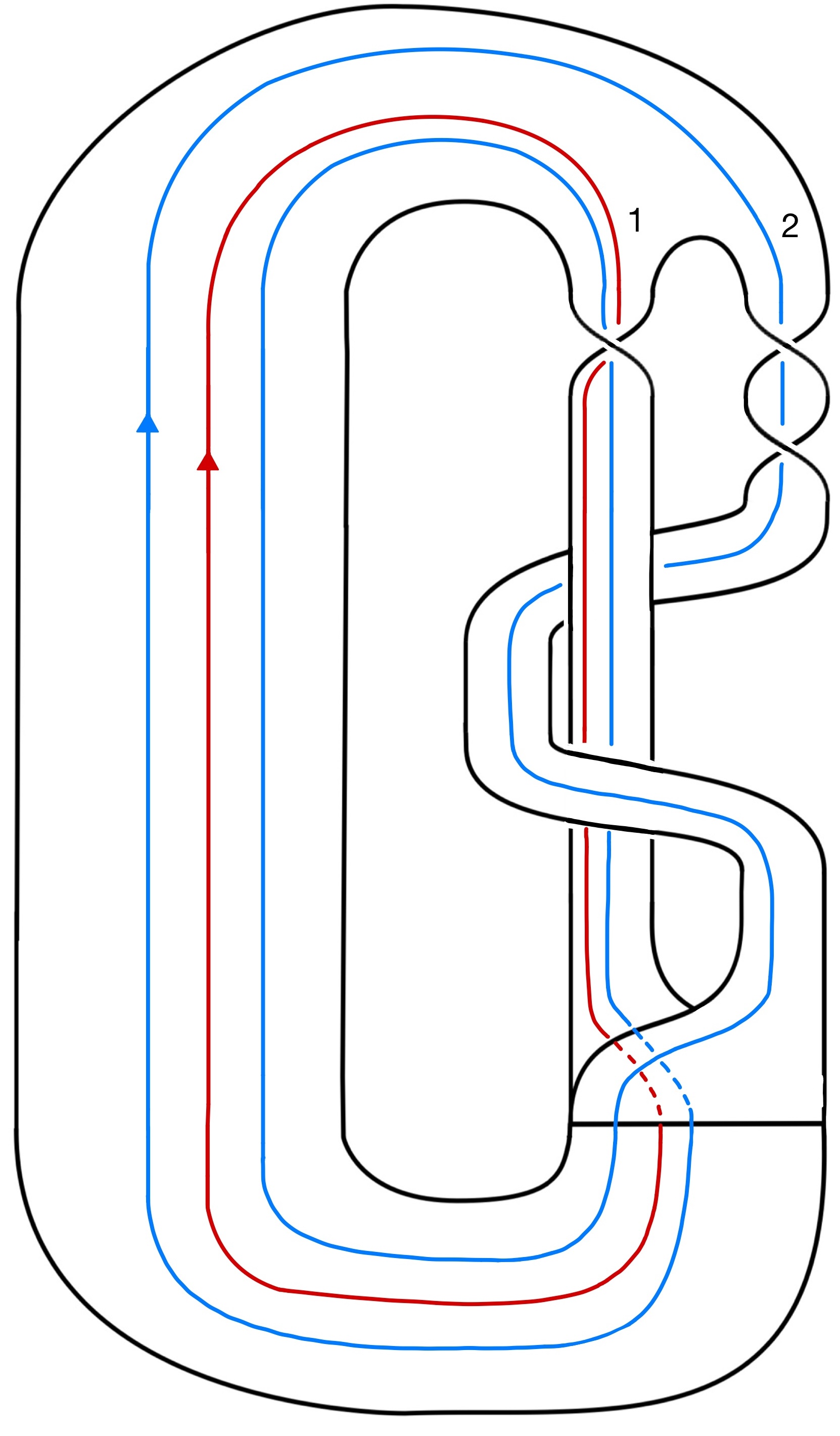}
    \caption{Template of KSE system with $L \approx 18.05$ extracted from linking numbers calculated on POD projections of UPOs. This represents the dynamics of the system in a topological manner, as trajectories move clockwise around this figure. All UPOs must follow the branched manifold with labels of the two branches yielding symbolic sequences for the orbits. Two low-period orbits used to identify this template are shown in red $(1)$ and blue $(21)$}\label{fig:templatePOD}
\end{figure} 
    
We use the POD described earlier to embed the dynamics of the Kuramoto-Sivashinsky PDE from the original phase space into an intermediary manifold having dimension three. Since the latent space is three-dimensional the Birman-Williams theorem can be applied and the procedure described earlier can be carried out to determine the template for the KSE. 
 
The simplified first-return map shown in figure \ref{fig:pod_return_map} has two monotonic branches and so the template has two branches labelled 1 and 2. With the five lowest period UPOs as input to the algorithm described in section \ref{sec:topology}, we obtain a unique candidate template with two branches. The candidate template is represented in figure \ref{fig:templatePOD} and is uniquely characterised by the following algebraic description: 
\begin{equation}\label{templatePOD}
(t_{1,1}, t_{1,2}, t_{2,2}, l_{1,2}) = (-1, -2, -2, 1).\end{equation}  The next ten orbits are incorporated into the input set, leading to the addition of 90 equations to the system (\ref{Formula}). All those are consistent with the template found, strongly suggesting that we have indeed identified the template of the chaotic attractor of the KSE PDE.

Note that after projecting the periodic orbits found in Section \ref{sec:finding_upos} onto the POD space we compute their linking numbers using the algorithm described by \citet{Qu_2021}. The linking numbers are given in table \ref{lknbr}. We did not calculate self-linking numbers for the periodic orbits, as this requires the choice of a framing, and it is not clear how to do this consistently given the dimensionality reduction. It is not necessary to compute self-linking numbers to determine a template, but without them it requires more periodic orbits to find a unique template\citep{Rosalie2016TemplatesAS}. 

Some of the UPOs considered have very long periods. We found that numerically calculating the linking numbers of these requires discretization of each loop using a very high number of points, given the necessity of accurately counting the crossings of periodic orbits in very close proximity. To resolve the most challenging cases, it was necessary to discretize the orbits with $2^{13}$ points.

\subsubsection{Autoencoder template}
\begin{figure}
\centering
\includegraphics[width=0.55\linewidth]{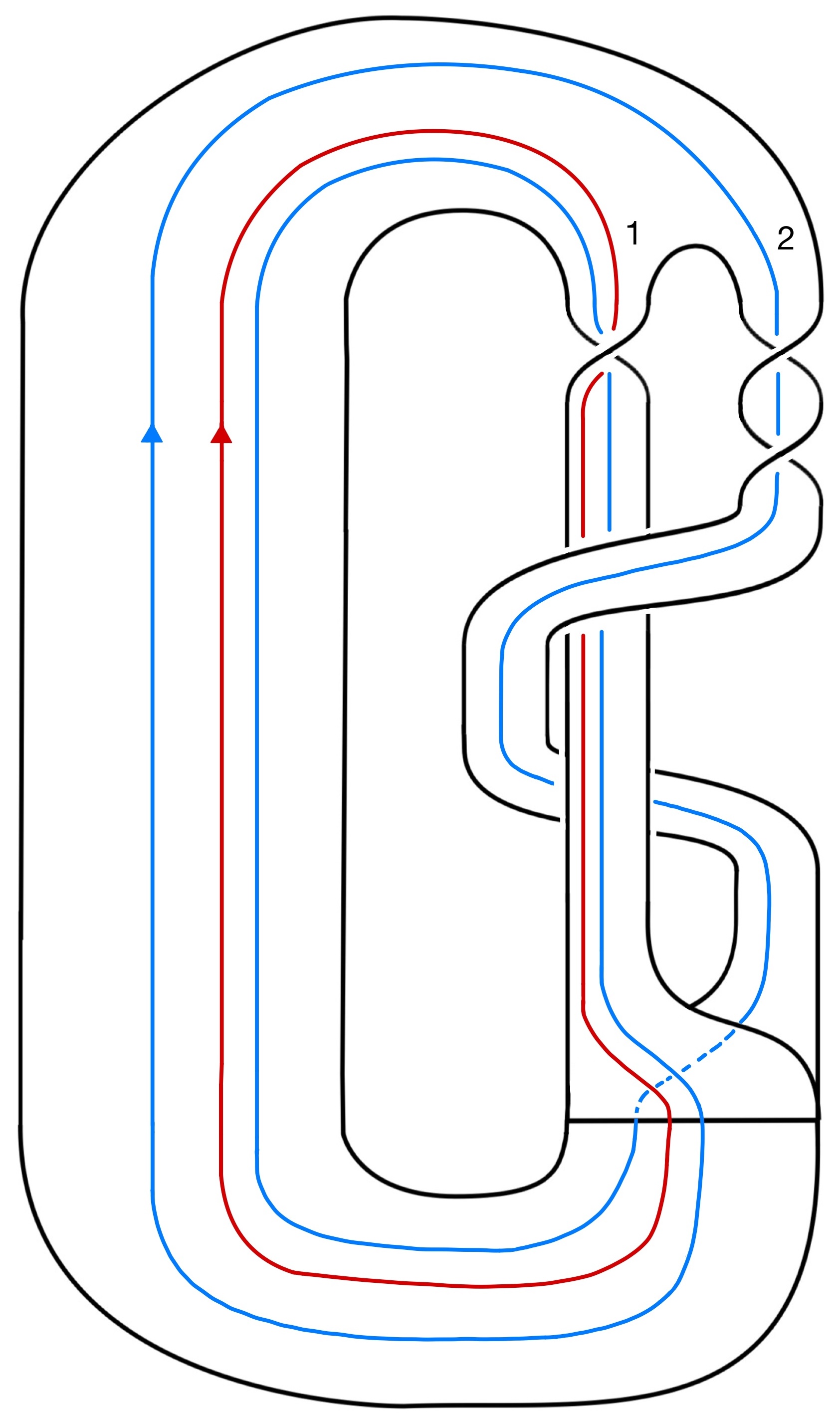}
\caption{As for figure \ref{fig:templatePOD} but with linking numbers calculated on autoencoder latent space images of UPOs.}
\label{fig:template}
\end{figure}
Using instead the images of the UPOs in the latent space of the autoencoder, the linking numbers were identical except for a global change of sign, i.e. all negative linking numbers in table \ref{lknbr} were replaced by positive numbers.
Again, the first-return map shown in figure  \ref{fig:first_return_map} has two monotonic branches and so the template has two branches labelled 1 and 2. With the five lowest period UPOs as input to the algorithm, we obtain a unique candidate template with two branches represented in figure \ref{fig:template} whose algebraic description is given by: \begin{equation}\label{template}(t_{1,1}, t_{1,2}, t_{2,2}, l_{1,2}) = (1, 2, 2, -1).\end{equation}
The next ten orbits are incorporated into the input set, leading to the addition of 90 equations to the system (\ref{Formula}). All those are consistent with the template found. The resulting template is the mirror-image of the one obtained with the POD method. By taking the template associated with the POD represented in figure \ref{fig:templatePOD}, we can flip it (so that the ``bottom'' becomes the ``front''). Then, by taking the mirror image, we obtain the template associated with the autoencoder in figure \ref{fig:template}. The two templates are diffeomorphic, but not isotopic\citep{MirrorTemplates}. The topological organisation of the periodic orbits is preserved up to a global change of sign. Consequently, the characterisation is independent, as intended, of the different coordinates resulting from different dimensionality-reduction techniques.

\subsection{Symbolic sequences}
A template provides a natural way to enumerate the periodic orbits: to each periodic orbit, one assigns a word whose letters represent the successive branches traversed. A collection of periodic orbits with unique symbolic sequences are a key element of a method for constructing a generating partition by interpolating the encoding map \citep{Plumecoq1, Plumecoq2}.

During the template identification we obtained several sets of possible names so, according to the algorithm described in section IV.A, we should add more orbits in the input set. However, in our particular case of the KSE, we are able to determine the symbolic sequences of the input set of periodic orbits using the POD return map as shown in figure \ref{fig:pod_first_return_map}. The set of symbolic sequences for the input set of orbits is presented in \cref{tab:UPOs}. The choice of the starting location of a symbolic sequence for a given orbit is arbitrary; we follow the convention given by \citet{hao1998applied}, i.e. choosing the sequence whose initial string has even parity. 


To check the validity of this set of symbolic sequences, we compute the expression of the linking numbers in terms of layering and torsion coefficients with respect to the symbolic sequences of the periodic orbits (right side in (\ref{Formula})). Then, we substitute the coefficients with the template identification found previously (\ref{templatePOD}). In all cases, there is agreement with the linking numbers obtained from the embedded periodic orbits in the POD space in table \ref{lknbr} (left handside in (\ref{Formula})). For instance, consider \begin{align*}
    Lk(211121, 21111212121)&=
l_{12}(\pi_{1} - \pi_{2} + 12)\\ &+ 14t_{11} + 15t_{12} + 4t_{22}\\
Lk(2111, 21121)&=\frac{l_{12}}{2}(-4\pi_1\pi_2 + \pi_1 + 2\pi_2 + 6)\\&+ \frac{9}{2}t_{11}+ \frac{9}{2}t_{12} + t_{22}.
\end{align*}
In the expressions above $\pi_i$ is the parity of branch $i$. By substituting with the identification (\ref{templatePOD}) we recover the linking numbers from \cref{lknbr}: 
\begin{align*}
    Lk(211121, 21111212121)&=-39=Lk(6_2,11_2)\\
Lk(2111, 21121)&=-12=Lk(4_1,5_2).
\end{align*}

We proceed similarly with the autoencoder method. Again, as shown in figure \ref{fig:first_return_map}, from the autoencoder return map we obtain the same set of symbolic words for the input set of periodic orbits. The linking numbers obtained from the symbolic sequences and the template identification (\ref{template}) are consistent with the linking numbers obtained from the embedded periodic orbits in the autoencoder latent space.

Thus consistent symbolic sequences for low-period UPOs are derived from two different dimensionality reduction methods and the symbolic sequences are validated by the template identification.

\section{Conclusion}
\label{sec:conclusion}

In this paper, we have found a candidate template for the topology of a chaotic attractor in the Kuramoto-Sivashinski equation. To our knowledge, this is the first attempt to do this in a PDE. While it is possible that our template would be shown to be invalid when other periodic orbits are included, it was found using only five UPOs and validated with a further ten, so we are confident it gives an accurate representation of the topology of the dynamics. Furthermore, equivalent topologies were found using two different dimensionality-reduction techniques: proper orthogonal decomposition and an autoencoder. Though neither method is guaranteed to preserve the topology in general \citep{Mindlin1}, the agreement between the two increases our confidence of the accuracy of the topology, and furthermore provides strong evidence that these dimensionality reduction methods are sufficiently accurate to capture the behaviour of the system in this case.

The main result of this paper is figure \ref{fig:templatePOD} (or equivalently figure \ref{fig:template}). We now have a qualitative interpretation of the dynamics: as trajectories move around the chaotic attractor, they are divided into two branches, which intertwine and eventually merge.
The bounding torus for this attractor has one hole, unlike the famous Lorenz attractor, but like the classical ODE systems of the R\"ossler and Sprott D attractors \citep{Letellier_2022}.
As with the templates for those latter ODE systems, the KSE attractor exhibits two branches, one of which has an additional twist, which demonstrates stretching-and-folding chaos (as opposed to `tearing-and-squeezing', the other standard mechanism). 

Topological descriptions of chaotic dynamics give a way of quantifying the stretching and folding processes which give rise to chaos. In a three-dimensional system like the R\"ossler system, this folding is directly visible from looking at the state space of the attractor. In higher-dimensions, projections and dimensionality reduction methods do not immediately provide this information, but a diagram of the template, as given in figures \ref{fig:templatePOD} and \ref{fig:template}, provide an immediate and intuitive interpretation of the dynamics.
A topological description of a chaotic dissipative PDE is particularly useful, as it can be used to compare between different dimensionality reduction techniques, and different discretizations of the system, to confirm that the essential dynamics are preserved, even if the exact quantitive details of the system may have been distorted.

The particular system and parameters studied here were chosen so that this work would be possible, and it is not directly generalisable to other dissipative PDEs: we stress that not only is an attractor with fractal dimension less than three necessary, we also must find an embedding of this into three dimensional Euclidean space. Indeed, in this case it was proven that a symbolic dynamics exists, which gave further expectation that our approach would be successful.

However, hope for applying topological methods to other PDEs is given by more modern techniques: assigning an $n$-dimensional cell-complex to a cloud of points on the attractor allows one to calculate invariants such homology groups, Betti number and Euler characteristic which provide information about the topological structure \citep{MULDOON19931, Reflexion4dim, ChaosHumSpeech, UnVeilTopChaos}. BraMAH (branched manifold analysis through homologies) can be applied to identify torsions, branch crossings and weak boundaries \citep{Charó_Artana_Sciamarella_2021, Char2020Noisedriven}. Recent work has improved this through the notion of templex \citep{Char2022TemplexAB} by endowing it with an oriented graph encoding the direction of the flow. A templex associated with a four-dimensional system is constructed in \citet{Char2022TemplexAB}. Though unlikely to be successful in the full state space of the discretized PDE, given the computational complexity of such a high-dimensional system, this is certainly possible in 4- or 5-dimensions, and so our approach of using an autoencoder to reduce the dimension can be combined with templexes to study the topology of more complicated PDE systems.

\begin{acknowledgments}
The authors are particularly indebted to an anonymous referee for finding an error in an initial version of the graphical representation of the template at the heart of this paper.
The authors wish to thank Omid Ashtari for fruitful discussions. This work was supported by the European Research Council (ERC) under the European Union's Horizon 2020 research and innovation programme (grant no. 865677).
\end{acknowledgments}

\bibliography{bib}

\end{document}